 \documentclass[10pt]{article} 
\newtheorem{defn}{D\'efinition}

\newtheorem{thm}{Theorem}
\newtheorem{lem}{Lemma}
\newtheorem{rem}{Remark}

\title{The Road Coloring Problem}
\date{Israel Journal of Mathematics}
\author{A.N. Trahtman\thanks{Bar-Ilan Univ., Math. Dep., 52900, Ramat Gan, Israel, Email: trakht@macs.biu.ac.il}
\thanks{http://www.cs.biu.ac.il/$\sim$trakht} 
} 
\begin{document}
\maketitle
\begin{abstract}
A synchronizing word of a deterministic automaton is a word in
the alphabet of colors (considered as letters) of its edges that
maps the automaton to a single state.  A coloring of edges of a
directed graph is synchronizing if the coloring turns the graph
into a deterministic finite automaton possessing a synchronizing word.

The road coloring problem is the problem of synchronizing coloring
of a directed finite strongly connected graph with constant outdegree
 of all its vertices if the greatest common divisor of lengths
of all its cycles is one. The problem was posed by Adler, Goodwyn
and Weiss over 30 years ago and evoked noticeable interest among
the specialists in the theory of graphs, deterministic automata
and symbolic dynamics.

The positive solution of the road coloring problem is presented.
\end{abstract}
 \section*{Introduction}
  The road coloring problem originates in \cite{AW} and was stated explicitly in
\cite {AGW} for a strongly connected
directed finite graph with constant outdegree of all its vertices where
the greatest common divisor (gcd) of lengths of all its cycles is one.
The edges of the graph are unlabeled.
The task is to find a labelling of the edges that turns the graph
into a deterministic finite automaton possessing a synchronizing word.
So the road coloring problem is connected with the problem of existence
of synchronizing word for deterministic complete finite automaton.

The condition on gcd is necessary \cite{AGW}, \cite{CKK}.
 It can be replaced by the equivalent property that there does not exist a
partition of the set of vertices on subsets $V_1$, $V_2$, ...,
 $V_k = V_1$ ($k>1$) such that every edge which begins in $V_i$ has its end
in $V_{i+1}$ \cite {CKK}, \cite {OB}.
The outdegree of the vertex can be considered also as the size of an
alphabet where the letters denote colors.

The road coloring problem is important in automata theory: a
synchronizing coloring makes the behavior of an automaton
resistant against input errors since, after detection of an error,
a synchronizing word can reset the automaton back to its original
state, as if no error had occurred. The problem appeared first in
the context of symbolic dynamics and is important also in this area.

Together with the \v{C}erny conjecture \cite{Pi}, \cite {Tt}, the
road coloring problem belongs to the most fascinating problems in
the theory of finite automata. The problem is discussed even in
"Wikipedia" - the popular Internet Encyclopedia.
However, at the same time it was considered as a "notorious open
problem" \cite{LM} and "unfeasible"  \cite{HJ}.

For some positive results in this area see \cite {BM},
\cite {Ca}, \cite {Fi}, \cite {GKS}, \cite {HJ}, \cite {JS}, \cite {Ka},
\cite {OB}, \cite {PS}; a detailed history of investigations
can be found in \cite {Ca}.

The concept from \cite{Fi} of the weight
of a vertex and the concept of a stable pair of states from
\cite{CKK}, \cite{Ka} with corresponding results and consequences
are used below.

We prove that the road coloring problem has a positive solution.
So a finite directed strongly connected graph with constant
outdegree of all vertices and with gcd of the lengths of all
 its cycles equal to one has a synchronizing coloring.
 \section*{Preliminaries}
A finite directed strongly connected graph with constant
outdegree of all its vertices where the gcd of lengths of all
 its cycles is one will be called {\it $AGW$ graph} as aroused by Adler,
Goodwyn and Weiss.

The bold letters will denote the vertices of a graph and the states
of an automaton.

If there exists a path in an automaton from the state $\bf p$ to
the state $\bf q$ and the edges of the path are consecutively
labeled by $\sigma_1, ..., \sigma_k$, then for
$s=\sigma_1...\sigma_k \in \Sigma^+$ let us write  ${\bf q}={\bf
p}s$.

Let $Ps$ be the map of the subset $P$ of states of an automaton by
help of  $s \in \Sigma^+$ and let $Ps^{-1}$ be the maximal set of
states $Q$ such that $Qs \subseteq P$. For the transition graph
$\Gamma$ of an automaton let $\Gamma s$
denote the map of the set of states of the automaton.

 $|P|$ - the size of the subset $P$ of states from an automaton
  (of vertices from a graph).

 A word $s \in \Sigma^+ $ is called a {\it synchronizing}
word of the automaton with transition graph $\Gamma$
if $|\Gamma s|=1$.

 A coloring of a directed finite graph is {\it synchronizing} if the
coloring turns the graph into a deterministic finite automaton
possessing a synchronizing word.

  A pair of distinct states $\bf p, q$ of an automaton (of vertices of
  the transition graph)  will be called  {\it synchronizing} if
  ${\bf p}s = {\bf q}s$ for some $s \in \Sigma^+$.
In the opposite case, if for any $s$ ${\bf p}s \neq {\bf q}s$, we
call the pair {\it deadlock}.

A synchronizing pair of states $\bf p$, $\bf q$ of an automaton is
called {\it stable} if for any word $u$ the pair ${\bf p}u, {\bf
q}u$ is also synchronizing \cite{CKK}, \cite{Ka}.

We call the set of all outgoing edges of a vertex a {\it bunch} if
 all these edges are incoming edges of only one vertex.

Let $u$ be a left eigenvector with positive components having no
common divisor of adjacency matrix of a graph with vertices
${\bf p}_1$, ..., ${\bf p}_n$.
The i-th component $u_i$ of the vector $u$ is called
{\it the weight} of the vertex ${\bf p}_i$ and denoted by
$w({\bf p}_i)$. The sum of the weights of the vertices from a
set $D$ is denoted by $w(D)$ and is called {\it the weight} of
 $D$ \cite{Fi}.

The subset $D$ of states of an automaton (of vertices of the
transition graph $\Gamma$ of the automaton) such that $w(D)$ is
maximal and $|Ds|=1$ for some word $s \in \Sigma^+$ let us call
{\it $F$-maximal} as introduced by Friedman \cite{Fi}.

The subset $\Gamma s$ of states (of vertices of the transition
graph $\Gamma$) for some word $s$ such that every pair of states
from the set is deadlock will be called an {\it $F$-clique}.
\section{Some properties of $F$-clique and of coloring free of stable pairs}
The road coloring problem was formulated for $AGW$ graphs \cite{AGW} and
only such graphs are considered below.
  We exclude from the consideration also the primitive cases
of graphs with loops and of only one color \cite{AGW}, \cite{OB}.
Let us formulate two important results from \cite{Fi} and \cite{Ka}
in the following form:
\begin{thm}  $\label {f0}$ \cite{Fi}
There exists a partition of $\Gamma$ on $F$-maximal sets (of the same weight).
 \end{thm}
Some side conditions on the $AGW$ graph stated in \cite{Fi} were
not used in the proof of this statement.
\begin{thm}  $\label {ck}$ \cite{Ka}
Let us consider a coloring of $AGW$ graph $\Gamma$. Stability of states
is a binary relation on the set of states of the obtained automaton;
denote this relation by $\rho$.

Then $\rho$ is a congruence relation, $\Gamma/\rho$ presents an
$AGW$ graph and synchronizing coloring of $\Gamma/\rho$ implies
synchronizing recoloring of $\Gamma$.
 \end{thm}
 The last theorem shows that if every $AGW$ graph has a
coloring with a stable pair, than every $AGW$ graph has
a synchronizing coloring.
\begin{lem}  $\label {f2}$
Let $w$ be the weight of $F$-maximal set of the $AGW$ graph
$\Gamma$ via some coloring.
  Then the size of every $F$-clique of the coloring is the same
and equal to $w(\Gamma)/w$ (the size of partition of $\Gamma$
on $F$-maximal sets).
 \end{lem}
 Proof. Two states from an $F$-clique could not belong to one
$F$-maximal set because this pair is not synchronizing.
By Theorem \ref{f0} there exists a partition of $\Gamma$ on
$F$-maximal sets of weight $w$. So the partition consists from
$w(\Gamma)/w$ $F$-maximal sets and to every $F$-maximal set belongs
at most one state from $F$-clique. Consequently, the size of
any $F$-clique is not greater than $w(\Gamma)/w$.

Let $\Gamma s$ be an $F$-clique.
 The sum of the weights ${\bf q}s^{-1}$ for all ${\bf q} \in \Gamma s$
 is the weight of $\Gamma$. So $$w(\Gamma)= \sum_{q \in \Gamma s}w({\bf
q}s^{-1})$$
 The number of addends (the size of the $F$-clique) is not greater
than $w(\Gamma)/w$. The weight of the set ${\bf q}s^{-1}$ for
every ${\bf q} \in \Gamma s$ is not greater than $w$. Therefore
${\bf q}s^{-1}$ is an $F$-maximal set of weight $w$ for every
${\bf q} \in \Gamma s$ and the size of any $F$-clique is
$w(\Gamma)/w$, the number of $F$-maximal sets in the corresponding
partition of $\Gamma$.
     \begin{lem}  $\label {f3}$
Let $F$ be  $F$-clique via some coloring of $AGW$ graph $\Gamma$.
For any word $s$ the set $Fs$ is also an $F$-clique and any state
[vertex] $\bf p$ belongs to some $F$-clique.
 \end{lem}
 Proof. Any pair $\bf p$, $\bf q$ from an $F$-clique $F$ is a deadlock.
To be deadlock is a stable binary relation, therefore for any word $s$
the pair ${\bf p}s$, ${\bf q}s$ from $Fs$ also is a deadlock.
So all pairs from $Fs$ are deadlocks.

For the $F$-clique $F$ there exists a word $t$ such that $\Gamma t = F$.
Thus $\Gamma ts = Fs$, whence $Fs$ is an $F$-clique.

For any ${\bf r}$ from a strongly connected graph
$\Gamma$, there exists a word u such that ${\bf r}={\bf p}u$
for $\bf p$ from the $F$-clique $F$,
whence ${\bf r}$ belongs to the $F$-clique $Fu$.
 \begin{lem}  $\label {f6}$
Let $A$ and $B$ ($|A|>1$) be distinct $F$-cliques via
some coloring  without stable pairs of the $AGW$ graph $\Gamma$.

 Then $|A| -|A \cap B| =|B| -|A \cap B| >1$.
 \end{lem}
Proof. Let us assume the contrary: $|A| -|A \cap B|=1$.
By Lemma \ref{f2}, $|A|=|B|$. So $|B| -|A \cap B|=1$, too.
 The pair of states ${\bf p} \in A \setminus B$ and
${\bf q} \in B \setminus A$ is not stable. Therefore for some word
$s$ the pair $({\bf p}s, {\bf q}s)$ is a deadlock.
Any pair of states from the $F$-clique $A$ and from the $F$-clique $B$
as well as from $F$-cliques $As$ and $Bs$ is a deadlock. So any pair
of states from the set $(A \cup B)s$ is a deadlock.
One has $|(A \cup B)s| = |A| + 1 > |A|$.

In view of Theorem \ref{f0}, there exists a partition of size $|A|$
(Lemma \ref{f2}) of $\Gamma$ on $F$-maximal sets. To every $F$-maximal set belongs
at most one state from $(A \cup B)s$ because every pair of states from this set is
a deadlock and no deadlock could belong to an $F$-maximal set.
This contradicts the fact that the size of $(A \cup B)s$ is greater than $|A|$.
\begin{lem}  $\label {f7}$
Let some vertex of $AGW$ graph $\Gamma$ have two incoming bunches.

 Then any coloring of $\Gamma$ has a stable couple.
 \end{lem}
Proof. If a vertex ${\bf p}$ has two incoming bunches from vertices ${\bf
q}$ and ${\bf r}$, then the couple ${\bf q}$, ${\bf r}$ is stable
for any coloring because ${\bf q}\alpha ={\bf r}\alpha =\bf p$ for
any letter (color) $\alpha \in \Sigma$.
\section{The spanning subgraph of cycles and trees with maximal number of edges
in the cycles}
\begin{defn}  $\label {d1}$
Let us call a subgraph $S$ of the $AGW$ graph $\Gamma$ a {\emph spanning
subgraph} of $\Gamma$ if to $S$ belong all vertices of $\Gamma$ and exactly one
outgoing edge of every vertex.

 A maximal subtree of the spanning subgraph $S$ with root on a cycle from $S$ and
having no common edges with cycles from $S$ is called a {\emph tree} of $S$.

 The length of path from a vertex ${\bf p}$ through the edges of the tree of the spanning
set $S$ to the root of the tree is called the {\emph  level} of ${\bf p}$ in $S$.
\end{defn}
\begin{rem}  Any spanning subgraph $S$ consists of disjoint cycles
and trees with roots on cycles; any tree and cycle of $S$ is defined
identically, the level of the vertex from cycle is zero,
the vertices of trees except root have positive level, the
vertex of maximal positive level has no incoming edge from $S$.
\end{rem}
  \begin{lem}  $\label {p1}$
Let $N$ be a set of vertices of level $n$ from some tree of the
spanning subgraph $S$ of $AGW$ graph $\Gamma$.

Then in a coloring of $\Gamma$ where all edges of $S$ have
the same color $\alpha$, any $F$-clique $F$ satisfies $|F \cap N| \leq 1$.
 \end{lem}
Proof. Some power of $\alpha$ synchronizes all states of given
level of the tree and maps them into the root. Any couple of
states from an $F$-clique could not be synchronized and therefore
could not belong to $N$.
\begin{lem}  $\label {f8}$
Let $AGW$ graph $\Gamma$ have a spanning subgraph $R$ of only disjoint cycles
(without trees).

 Then $\Gamma$ also has another spanning subgraph with exactly one vertex of
maximal positive level.
\end{lem}
Proof.
The spanning subgraph $R$ has only cycles and therefore the levels of all vertices
are equal to zero.
In view of gcd =1 in the strongly connected graph $\Gamma$, not all edges
belong to a bunch. Therefore there exist two edges
$u ={\bf p} \to {\bf q} \not\in R$ and $v ={\bf p} \to {\bf s} \in R$
with common first vertex ${\bf p}$ but such that ${\bf q} \neq {\bf s}$.
Let us replace the edge $v={\bf p} \to {\bf s}$ from $R$ by $u$.
Then only the vertex ${\bf s}$ has maximal level $L>0$ in the new spanning subgraph.
 \begin{lem}  $\label {f9}$
Let any vertex of an $AGW$ graph $\Gamma$ have no two incoming bunches.

Then $\Gamma$ has a spanning subgraph such that all its vertices
of maximal positive level belong to one non-trivial tree.
 \end{lem}
Proof.
Let us consider a spanning subgraph $R$ with a maximal number of vertices [edges]
in its cycles. In view of Lemma \ref{f8}, suppose that $R$ has
non-trivial trees and let $L>0$ be the maximal value of the level of a vertex.

Further consideration is necessary only if at least two
vertices of level $L$ belong to distinct trees of $R$ with
distinct roots.

Let us consider a tree $T$ from $R$ with vertex ${\bf p}$ of maximal
level $L$ and edge $\bar{b}$ from vertex ${\bf b}$ to the tree root
${\bf r} \in T$ on the path of length $L$ from ${\bf p}$. Let the root
${\bf r}$ belong to the cycle $H$ of $R$ with the edge
$\bar{c}={\bf c} \to {\bf r} \in H$. There exists also an edge
$ \bar{a}={\bf a} \to {\bf p}$ that does not belong to $R$ because
$\Gamma$ is strongly connected and ${\bf p}$ has no incoming edge from $R$.

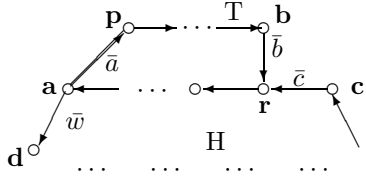
\begin{picture}(100,80)
\end{picture}
\begin{picture}(150,80)
\multiput(61,33)(26,0){3}{\circle{4}}
\put(87,54){\vector(0,-1){18}}
\put(87,56){\circle{4}}
\put(36,56){\circle{4}}
\put(0,10){\circle{4}}

\multiput(85,33)(26,0){2}{\vector(-1,0){21}}
\put(123,11){\vector(-1,2){10}}
\put(12,32){\vector(-1,-2){10}}

\put(13,33){\circle{4}}
\put(27,58){$\bf p$}
\put(85,24){$\bf r$}
\put(3,31){$\bf a$}
\put(-10,4){${\bf d}$}

\put(120,31){$\bf c$}
\put(91,58){$\bf b$}

\put(38,56){\vector(1,0){16}}
\put(69,56){\vector(1,0){17}}

\put(56,54){$\cdots$}
\put(39,30){$\cdots$}
\multiput(16,0)(28,0){4}{$\cdots$}
\put(32,33){\vector(-1,0){17}}
\put(27,40){$\bar{a}$}
\put(12,18){$\bar{w}$}
\put(98,34){$\bar{c}$}
\put(90,45){$\bar{b}$}
\put(65,10){H}
\put(72,59){T}
\put(14,34){\vector(1,1){20}}
\put(14,35){\line(1,1){20}}
 \end{picture}

Let us extend the path from ${\bf p}$ to ${\bf r}$ of maximal
length $L$ in $T$ by one or two in the three following ways:

1) replace the edge $\bar{w}$ from $R$ with first vertex ${\bf a}$
by the edge $\bar{a}={\bf a} \to {\bf p}$,

2) replace the edge $\bar{b}$ from $R$ by some other outgoing
edge of the vertex ${\bf b}$,

3) replace the edge $\bar{c}$ from $R$ by some other outgoing edge of
the vertex ${\bf c}$.

Our aim is to extend the maximal level of the vertex on the extension of the
tree $T$ much more than the maximal level of vertex of other
trees from $R$. Let us begin with

1)  Suppose first ${\bf a} \not\in H$. In this case either a new cycle is added
to $R$ extending the number of vertices in its cycles in spite of the choice of $R$
 or in the new spanning subgraph the level of ${\bf a}$ is $L+1$ and the vertex
${\bf r}$ is a root of the new tree containing all vertices of maximal level
(the vertex ${\bf a}$ or its ancestors in R).

So let us assume ${\bf a} \in H$ and suppose $\bar{w}={\bf a} \to {\bf d} \in H$.
In this case the vertices ${\bf p}$, ${\bf r}$ and ${\bf a}$ belong to
a cycle $H_1$ with new edge $\bar{a}$ of a new spanning subgraph $R_1$.
So we have the cycle $H_1 \in R_1$ instead of $H \in R$.
If the length of path from ${\bf r}$ to ${\bf a}$
in $H$ is $r_1$ then $H_1$ has length $L+r_1+1$. A path to ${\bf r}$
from the vertex ${\bf d}$ of the cycle $H$ remains in $R_1$.
Suppose its length is $r_2$. So the length of the cycle $H$
is $r_1+r_2+1$. The length of the cycle $H_1$ is not greater than
the length of $H$ because the spanning subgraph $R$ has maximal number of
edges in its cycles. So $r_1+r_2+1 \geq L+r_1+1$, whence $r_2
\geq L$. If $r_2 > L$, then the length $r_2$ of the path from
${\bf d}$ to $\bf r$ in a tree of $R_1$ (and the level of ${\bf d}$)
 is greater than $L$ and the level of ${\bf d}$ (or of some other ancestor of
${\bf r}$ in a tree from $R_1$) is the desired unique maximal
level.

 So assume for further consideration $L=r_2$ and
 ${\bf a} \in H$.  It holds in the case of adding of incoming
edge of any vertex of maximal level with root in $H$.

2) Suppose the set of outgoing edges of the vertex ${\bf b}$ is
not a bunch. So one can replace in $R$ the edge $\bar{b}$ from the
vertex ${\bf b}$ by an edge $\bar{v}$ from ${\bf b}$ to a vertex
${\bf v}\neq {\bf r}$.

The vertex ${\bf v}$ could not belong to $T$
because in this case a new cycle is added to $R$ and therefore
a new spanning subgraph has a number of vertices in the cycles greater than in $R$.

If the vertex ${\bf v}$ belongs to another tree of $R$ but not to cycle,
then $T$ is a part of a new tree $T_1$ with a new root of a new spanning
subgraph $R_1$ and the path from ${\bf p}$ to the new root is extended.
So only the tree $T_1$ has states of new maximal level.

If ${\bf v}$ belongs to some cycle $H_2 \neq H$ from $R$,  then
together with replacing $\bar{b}$ by $\bar{v}$, we replace also the
 edge $\bar{w}$ by $\bar{a}$. So we extend the path from ${\bf p}$
to the new root ${\bf v}$ at least by the edge
$\bar{a}={\bf a} \to {\bf p}$ and by almost all
edges of $H$. Therefore the new maximal level $L_1 > L$ has either
the vertex ${\bf d}$ or its ancestors from the old spanning subgraph
$R$.

Now it remains only the case when ${\bf v}$ belongs to the cycle
$H$. The vertex ${\bf p}$ also has level $L$ in new tree $T_1$
with root ${\bf v}$. The only difference between $T$ and $T_1$
(just as between $R$ and $R_1$) is the root and the incoming edge
of the root. The new spanning subgraph $R_1$ has also a maximal
number of vertices in cycles just as $R$. Let $r_3$ be the length
of the path from ${\bf d}$ to the new root ${\bf v} \in H$.

For the spanning subgraph $R_1$, one can obtain $L=r_3$ just as it
was done on the step 1) for $R$. From  ${\bf v} \neq {\bf r}$
follows $r_3 \neq r_2$, though $L=r_3$ and $L=r_2$.

So for further consideration suppose that the set of outgoing
edges of the vertex ${\bf b}$ is a bunch to ${\bf r}$.

3) The set of outgoing edges of the vertex ${\bf c}$ is
not a bunch to ${\bf r}$ because ${\bf r}$ has another bunch from ${\bf b}$.

Let us replace in $R$ the edge $\bar{c}$ by an edge $\bar{u}={\bf
c} \to {\bf u}$ such that ${\bf u}\neq {\bf r}$. The vertex ${\bf
u}$ could not belong to the tree $T$ because in this case the cycle $H$
is replaced by a cycle with all vertices from $H$ and some vertices of $T$
 whence its length is greater than $|H|$. Therefore the new spanning
subgraph has a number of vertices in its cycles greater than in
spanning subgraph $R$ in spite of the choice of $R$.

So remains the case ${\bf u} \not\in T$. Then the tree $T$ is a
part of a new tree with a new root and the path from ${\bf p}$ to
the new root is extended at least by a part of $H$ from the former
root $\bf r$. The new level of ${\bf p}$ therefore is maximal and
greater than the level of any vertex in some another tree.

Thus anyway there exists a spanning subgraph with vertices
of maximal level in one non-trivial tree.
\begin{thm}  $\label {t1}$
 Any $AGW$ graph $\Gamma$ has a coloring with stable couples.
 \end{thm}
Proof. By Lemma \ref{f7}, in the case of vertex with two incoming bunches
$\Gamma$ has a coloring with stable couples. In opposite case, by
 Lemma \ref{f9}, $\Gamma$ has a spanning subgraph $R$ such
that the vertices of maximal positive level $L$ belong to one
tree of $R$.

Let us give to the edges of $R$ the color $\alpha$
and denote by $C$ the set of all vertices from
the cycles of $R$.
Then let us color the remaining edges of $\Gamma$ by other colors
arbitrarily.

By Lemma \ref{f3}, in a strongly connected graph $\Gamma$ for
every word $s$ and $F$-clique $F$ of size $|F| > 1$, the set $Fs$
also is an $F$-clique (of the same size by Lemma \ref{f2}) and for
any state $\bf p$ there exists an $F$-clique $F$ such that ${\bf
p} \in F$.

In particular, some $F$ has non-empty intersection with the set
$N$ of vertices of maximal level $L$. The set $N$ belongs to one
tree, whence by Lemma \ref{p1} this intersection has only one
vertex.
 The word $\alpha^{L-1}$ maps $F$ on an $F$-clique $F_1$ of size
$|F|$. One has $|F_1 \setminus C|=1$ because the sequence of edges of
color $\alpha$ from any tree of $R$ leads to the root of
the tree, the root belongs to a cycle colored by $\alpha$ from $C$ and
 only for the set $N$ with vertices of maximal level holds
 $N\alpha^{L-1} \not\subseteq C$. So
 $|N\alpha^{L-1}  \cap F_1|=|F_1 \setminus C|=1$ and
$|C \cap F_1|=|F_1|-1$.

Let the integer $m$ be a common multiple of the lengths
of all considered cycles from $C$ colored by $\alpha$.
 So for any $\bf p$ from $C$ as well as from $F_1 \cap C$
holds ${\bf p}\alpha^m={\bf p}$. Therefore for an $F$-clique
$F_2=F_1\alpha^m$ holds $F_2 \subseteq C$ and
$C \cap F_1 =F_1 \cap F_2$.

Thus two $F$-cliques $F_1$ and $F_2$ of size $|F_1|>1$ have
$|F_1|-1$ common vertices. So $|F_1 \setminus (F_1 \cap F_2)|=1$.
Consequently, in view of Lemma \ref{f6}, there exists a stable
couple in the considered coloring.
\begin{thm}  $\label {t}$
Every $AGW$ graph $\Gamma$ has synchronizing coloring.
 \end{thm}
  The proof follows from Theorems \ref{t1} and \ref{ck}.

 \end{document}